\documentclass[aps,preprintnumbers,amsmath,amssymb,twocolumn, tightenlines,superscriptaddress,nofootinbib]{revtex4}
\usepackage{graphicx}
\usepackage{slashed}
\usepackage{lipsum}
\usepackage{tikz,mathpazo}
\usetikzlibrary{shapes.geometric, arrows}
\usepackage[none]{hyphenat}
\usepackage[english]{babel}
\usepackage{changes}
\newcommand{\cmmnt}[1]{}

\begin{document}

\title{Building imaginary-time thermal field theory with artificial neural networks}

\author {Tian Xu} 
\affiliation{Physics Department, Beihang University, 37 Xueyuan Rd, Beijing 100191, China}

\author{Lingxiao Wang}
\affiliation{Shanghai Research Center for Theoretical Nuclear Physics, National Natural Science Foundation of China and Fudan University, Shanghai 200438, China}
\affiliation{Frankfurt Institute for Advanced Studies, Ruth-Moufang-Str. 1, 60438 Frankfurt am Main, Germany}

\author{Lianyi He}
\affiliation{Department of Physics, Tsinghua University, Beijing 100084, China}

\author{Kai Zhou}
\affiliation{School of Science and Engineering, The Chinese University of Hong Kong, Shenzhen (CUHK-Shenzhen), Guangdong, 518172, China}
\affiliation{Frankfurt Institute for Advanced Studies, Ruth-Moufang-Str. 1, 60438 Frankfurt am Main, Germany}

\author {Yin Jiang} \email{jiang_y@buaa.edu.cn}
\affiliation{Physics Department, Beihang University, 37 Xueyuan Rd, Beijing 100191, China}
\date{\today}

\begin{abstract}

In this study, we introduce a novel approach in quantum field theories to estimate the action using the artificial neural networks (ANNs). The estimation is achieved by learning on system configurations governed by the Boltzmann factor, $e^{-S}$ at different temperatures within the imaginary time formalism of thermal field theory. We focus on 0+1 dimensional quantum field with kink/anti-kink configurations to demonstrate the feasibility of the method. The continuous-mixture autoregressive networks (CANs) enable the construction of accurate effective actions with tractable probability density estimation. Our numerical results demonstrate that this methodology not only facilitates the construction of effective actions at specified temperatures but also adeptly estimates the action at intermediate temperatures using data from both lower and higher temperature ensembles. This capability is especially valuable for the detailed exploration of phase diagrams.
\end{abstract}

\maketitle

\section{Introduction}

Lattice simulation is an important systematic method to solve strongly correlating and interacting systems in the framework of quantum field theory at finite temperatures \cite{Muroya:2003qs,Ratti:2018ksb}. By sampling numerous configurations according to the action, physical observables are computed by taking ensemble average. The background physical mechanism can only be explored by certain integral quantities because of the large amount of configurations and unavoidable fluctuations during the sampling process \cite{BMW:2008jgk,Ballini:2024wbe}. In order to test or realize a certain physical mechanism, people have developed different effective models. Such models are built with specific effective degrees of freedom (\textit{d.o.f}) and further introducing proper interactions between them \cite{Greensite:2011zz}. Starting from the fundamental theory, building an effective model is usually not so straightforward because the difficulty of choosing the key \textit{d.o.f} and including the corresponding interactions. There are several successful examples, such as the Cooper pair for superconductivity, vortex for the XY model and various soliton solutions for the quantum chromodynamics (QCD) \cite{Bardeen:1957mv,Kosterlitz:1973xp,Nambu:1974zg,Harrington:1978ve,Fritzsch:1990kw,Schafer:1996wv}. Most of such key \textit{d.o.f} are obtained by solving the semi-classical equations. To complete the effective model, the fluctuations about the chosen \textit{d.o.f} should be integrated out properly, which is usually a hard and tedious work. 

From the viewpoint of functional integration of quantum field theory, the Lagrangian density is encoded in the distribution of the configuration set \cite{altland:2010condensed}. The emerging probability of each configuration is determined by its action. If the correctly distributed ensemble is known, for example by coarse-graining the fundamental lattice configuration into the chosen effective one, the action can be extracted by estimating the probability of the configuration~\cite{Sonoda:2020vut}, which is almost impossible task \cite{Schaefer:2010hu,Pan:2022fgf} with traditional methods for such a high dimensional distribution. But the recent popular deep learning frameworks are good at solving such problems \cite{Carrasquilla:2016oun, Carleo:2017nvk, Carleo:2019ptp, Zhou:2023pti}. Basing on the variational ansatz that decomposing the probability of a lattice configuration into conditional probability product, a class of autogressive networks have developed for the probability estimation issues \cite{Wu:2019elz,Sharir:2020tbn,Luo:2023opo}. For classical systems, such frameworks have already been introduced to study underlying interaction details of condensed matter, chemistry and biology systems \cite{fujita:2007modeling, goldman:2024generating, Wang:2020hji, Wang:2020neural}. Although there have been some attempts to learn the action in quantum lattice field theories~\cite{Shanahan:2018vcv, Blucher:2020mjt, Favoni:2020reg}, research into estimating under external parameters, e.g., the temperature dependence, remains notably scarce.

The imaginary-time thermal field theory reformulates the quantum statistics via compacting the time direction onto a imaginary- time ring~\cite{Allen:1986qi}. As a result, the complex weight $\exp(i S)$ of a configuration becomes a real probability $\exp(-S)$. And if one discretizes the integral over imaginary time, it can be found that the temperature dependence of kinetic and potential part are different but explicit \cite{Bruckmann:2007ru, Lopez-Ruiz:2016tnf}. This makes almost the same procedure as the classical case can work again for quantum ones, except there are more than one ensemble is need to determine the whole phase diagram along temperature axis. In this work we will review the classical case briefly to clarify the paradigm of constructing the effective model with artificial neural networks (ANNs). And then the quantum version will be discussed. One will see that if the potential part is independent on the imaginary time, only two ensembles at different temperature are enough to determine the action of each configuration if the Lagrangian density is composed with the sum of kinetic and local potential terms. We will use a quantum mechanics example (0+1D field theory) to  show the suggested procedure and adopt the continuous-mixture autoregressive networks (CANs) to estimate the probability \cite{Wang:2020hji}. From the numerical experiment one will see that to predict the action of one sample at a certain temperature, only two different ensembles are enough. And in the interpolation case, i.e. the predicting temperature falls in the range of the two known temperature, works best. It sounds acceptable because when the phase structure is roughly characterized at two ends by a effective model, its estimation for intermediate states would not be too wrong.  

\section{Neural Networks for Classical Statistics}
In classical statistics the probability of one sample in a physical ensemble is determined by the energy of the sample, which is given by the Hamiltonian $H$ of the model. Taking a spin system for example, a configuration is represented as a set of spin values at every site as $\sigma=\{s_1, s_2, ..., s_N\}$, and the thermodynamic properties of the system at a certain temperature $T$ is governed by the ensemble whose samples are distributed according to the Hamiltonian $H(\sigma)$ as the Boltzmann distribution 
\begin{eqnarray}
P(\sigma)\propto\exp(-H(\sigma)/T).    
\end{eqnarray}
Once the Hamiltonian of the system is known one can obtain the ensemble by some certain sampling methods, such as the Markov chain Monte Carlo(MCMC). In this work we focus on the inverse problem, i.e. extracting the Hamiltonian, or estimating probability equivalently, when an ensemble $\{\sigma^{(1)}, \sigma^{(2)}, ..., \sigma^{(M)}\}$ is known, where $M$ is a large enough integer for an ensemble. Clearly, it is difficult to estimate the high-dimensional probability distribution with traditional method, i.e., the so-called curse of dimensionality~\cite{Carrasquilla:2016oun}. Fortunately the ANN provides a good solution to this problem \cite{Shanahan:2018vcv,Wang:2020hji}. Such a network could be helpful in two aspects. Firstly, a numerical effective model can be constructed via this network in a more straightforward way. Explicitly, basing the original ensemble each sample $\sigma^{(i)}$ can be reconstructed by an effective mode $\Sigma^{(i)}$, such as the vortex in XY model, to obtain a correctly distributed ensemble \cite{song:2023monte}. With the network the energy or equivalently probability of $\Sigma^{(i)}$ can be extracted, and thus a numerical interaction of the effective mode has been obtained 
\begin{eqnarray}
    H(\Sigma)=-T \ln (P(\Sigma))+C
\end{eqnarray} 
up to a constant $C$, which corresponds to the partition function. Secondly, if the ensemble can be measured in laboratory, even in a more macroscopic degree of freedom, an numerical Hamiltonian can be constructed by the network. Once a Hamiltonian has been obtained, the system properties at different temperatures can be estimated by generating a new ensemble according to the Boltzmann distribution with the help of a standard sampling method. It should be noted that the temperature dependence is explicitly introduced as a physical prior for the construction of the numerical Hamiltonian with ANNs~\cite{Wang:2020neural}.

\section{Neural Networks for Quantum Statistics}
\label{sec:quantum}

We further consider the quantum version of the above problem, i.e. basing on several known ensembles at one or more temperatures to extract the action by estimating the probability density of each sample. The quantum effects can be introduced into classical statistics by either considering the Hamiltonian as a quantum operator $\exp(-{\hat H}/T)$ or integrating over all the possible evolution processes which is known as the path integral approach of the quantum theory. At zero temperature, the weight of a quantum state is drawn as a complex factor, $\exp(-i S)$, which is inapplicable for real-valued ANNs. However, in finite temperature case the imaginary formalism compacts the time onto a ring with radius $\beta=T^{-1}$ of imaginary time $\tau=i t$. In such a formalism the partition function is 
\begin{eqnarray}
Z=\int D\Phi \exp(-S[\Phi]),
\end{eqnarray}
where the action is $S[\Phi]=\int_0^\beta d\tau d^3x[(\partial_\tau\Phi)^2+(\nabla\Phi)^2+V(\Phi)]$ in general if we only consider the usual local and Lorentz covariant Bosonic field system for example. And for fermionic cases a proper Hubbard-Stratonovich transformation can be applied to obtain a bonsonic one. Although the temperature-dependence appears more complicated than the classical case which is $e^{-E/T}$, it is still tractable if the discretization formalism is rewritten down explicitly as 
\begin{eqnarray}
S[\Phi]&&=\sum \Delta\tau (\Delta x)^3\left[(\frac{\Delta \Phi}{\Delta \tau})^2+(\nabla\Phi)^2+V(\Phi)\right]\nonumber\\
&&=\sum  (\Delta x)^3\left[\frac{(\Delta \Phi)^2}{\Delta \tau}+\Delta\tau((\nabla\Phi)^2+V(\Phi))\right]\nonumber\\
&&=\beta^{-1}K+\beta V
\end{eqnarray}
where $\Delta\tau=\beta/N_\tau$, { the first term is part of kinetic term denoted as $K\equiv N_\tau\sum  (\Delta x)^3{(\Delta \Phi)^2}$, and the second term includes all the time-independent terms denoted as $V\equiv N_\tau^{-1}\sum  (\Delta x)^3{[(\nabla\Phi)^2+V(\Phi)]}$. The temperature dependence of these two terms are different and is separable with regard to the quantum fields.} It is evident that once actions of a sample at two given temperatures can be estimated as
\begin{eqnarray}
&&S_1[\Phi]=\beta_1^{-1}K[\Phi]+\beta_1 V[\Phi]+C_1\nonumber\\
&&S_2[\Phi]=\beta_2^{-1}K[\Phi]+\beta_2 V[\Phi]+C_2,
\end{eqnarray}
the $K$ and $V$ term can be solved out and the action at any third temperature is
\begin{eqnarray}
\label{interp}
S_3[\Phi]=\frac{\beta_1(\beta_3^2-\beta_2^2)}{\beta_3(\beta_1^2-\beta_2^2)}S_1+  \frac{\beta_2(\beta_1^2-\beta_3^2)}{\beta_3(\beta_1^2-\beta_2^2)}S_2+C_3,
\end{eqnarray}
where 
\begin{eqnarray}
C_3=\frac{\beta_1(\beta_2^2-\beta_3^2)}{\beta_3(\beta_1^2-\beta_2^2)}C_1+  \frac{\beta_2(\beta_3^2-\beta_1^2)}{\beta_3(\beta_1^2-\beta_2^2)}C_2.
\end{eqnarray}
Employing the ANNs as the classical case twice, we can extract the probability(up to a global constant) of any sample in a correctly distributed ensemble at two given temperatures. Equivalently, the temperature-independent terms $K$ and $V$ can be determined by feeding two ensembles—-generated via certain sampling algorithms-—into ANNs. The network is then trained to output the numerical action or probability density associated with each sample. Utilizing two such ensembles at distinct temperatures allows for the encoding of interaction details within two respective ANNs, denoted $T_1$ and $T_2$. This methodology obviates the need for an analytical expression of the Lagrangian density.

\begin{figure}[!hbtp]
\centering
\includegraphics[width=0.5\textwidth]{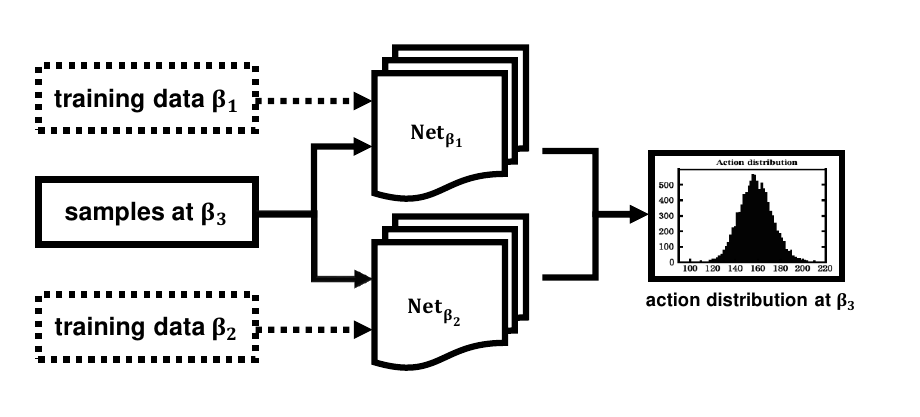}
\caption{\label{fig:fig2}Schematic of neural network predictions for the quantum statistics.}
\end{figure}
A schematic flowchart is shown in Fig.~\ref{fig:fig2}. This quantum version algorithm indicates only two ensembles are required to reconstruct the interaction detail of a system, although the whole evolution history along the imaginary time will contribute to the quantum partition function in principle. 

\section{0+1D quantum field model and kinks}
A simple 1D quantum mechanical system is enough for the experiment of the above algorithm. The Lagrangian is
\begin{eqnarray}
L=\frac{1}{2}\left(\frac{dx}{d\tau}\right)^{2}+V_{k}(x).
\end{eqnarray}
We introduce a standard double-well potential as,
\begin{eqnarray}
V_{k}(x)=\frac{\lambda_{k}}{4}\left(x^{2}-\frac{\mu_{k}^{2}}{2 \lambda_k}\right)^{2}.
\end{eqnarray}
Here $\lambda_k$ is the coupling constant with mass dimension $[\lambda_k] = [m]^5$ and a parameter $\mu_k$ with mass dimension $[\mu_k] = [m]^{3/2}$. Suppose the particle mass parameter (in the kinetic energy term) is m, we will then work with dimensionless quantities rescaled by proper powers of m (e.g., $\mu^2_k/m^3\rightarrow\mu^2_k$ and $\lambda_k/m^5\rightarrow\lambda_k$) throughout this paper.
The properties of the system at finite temperature $T$ can be described through the partition function
\begin{eqnarray}
Z &&= \int _{x(\beta)=x(0)}Dx\ e^{-S_E[x(\tau)]}\nonumber\\
&&=\int\prod_{j=-N+1}^{N+1}\frac{dx_{j}}{\sqrt{2\pi a}}\nonumber\\
&&\times\exp\left\{-\sum_{i=-N+1}^{N+1}\left[\frac{(x_{i+1}-x_{i})^{2}}{2a}+a{V_k(x_{i})}\right]\right\},
\end{eqnarray}
where $k_B = \hbar = 1$, $\beta = 1/T$ is the inverse of temperature T, and $S_E$ the Euclidean action in the imaginary formalism given by

\begin{eqnarray}
S_{E}[x(\tau)]&&=\int_{0}^{\beta}d\tau\ \mathcal{L}_{E}[x(\tau)]\nonumber\\
&&=\int_{0}^{\beta}d\tau\ \left[\frac{1}{2}\left(\frac{dx}{d\tau}\right)^{2}+V_{k}(x)\right] .
\end{eqnarray}

When the interaction coupling is not too large, the system is governed by the semi-classical solutions of the equation of motion. In order to derive a non-trivial semi-classical solution in the 0+1-dimensional field system explicitly, we can consider the 1D quantum mechanical system with Higgs-like interaction potential. This is a very popular potential to realize the spontaneous symmetry breaking~\cite{tHooft:1999cgx,Bruckmann:2007ru,Chen:2022ytr} in the higher dimensional system, while for the 1D case no such a mechanism exist. Instead, for such a system a type of tunneling solution known as the kink is a close analog of the instanton in QCD. 

The semi-classical solution is obtained by minimizing the action and solving the equation of motion. There are also two nontrivial solutions given by
\begin{eqnarray}
x(\tau) = \pm\frac{\mu_k}{\sqrt{\lambda_k}} \tanh  \left[\frac{\mu_{k}}{\sqrt{2}}(\tau -\tau_0)\right].
\end{eqnarray}
These solutions approach $\pm \mu_k/\sqrt{\lambda_k}$ at $\tau=\pm \infty$. This means these two solutions interpolate between two minima over inﬁnite long imaginary time. Such a behavior is consistent with the ground state of the Schrodinger equation, which is supposed to peak at the middle of the two minima of the potential. The above solutions with plus and minus sign are respectively called the kink and anti-kink solutions which represent the tunneling process if one transforms the solution back to the real-time formulation.

In the numerical simulation, as suggested in Ref.~\cite{Schafer:2004xa}, we choose $\lambda_k=4$ and $\mu_{k}/\sqrt{\lambda_{k}}=1.4$ and adopt the traditional MCMC for sampling at different temperatures \cite{Schafer:2004xa}. We will try to used a ANN to estimate the probability of each sample and to reconstruct the Lagrangian density using two ensembles.  In order to apply the MCMC sampling the continuous $x(\tau)$ should be firstly discretized, and smaller lattice size certainly requires larger number of steps before convergence is achieved. As a practical example in our work, the simulations for different temperatures ($\beta = T^{-1} = 80,40,20$) were done with the  same number of lattice size(256) and sweeps ranging  ($N_{MC} = 5 \times 10^6$). After discarding the first 50,000 steps we chose 10,000 configurations randomly in the following sequence as training set and another 10,000 samples as testing set for each $\beta$. 

\section{Continuous-mixture Autoregressive Network(CAN) to build actions}
CAN is a suitable neural network for extracting the probability density for each sample with continuous \textit{d.o.f} \cite{Wu:2019elz,Wang:2020hji}. Here we adopt it to learn the probability density of each sample in quantum systems. The algorithm is constructed according to the Maximum Likelihood Estimate (MLE), which is employed to estimate the probability density in an unsupervised manner \cite{pan:2002maximum}. Two basic properties of probability should be satisfied. Firstly, the given probability is positive. Secondly, two similar configuration gives similar and continuous values of probability. To achieve the two properties, in the CAN, we propose factorizing the whole probability of a sample as the product of conditional probabilities at each site and using an appropriate mixture of Beta distributions as the prior probability to ensure positive and continuous requirements. The Beta distribution ${\mathcal B}(a, b)$ with two parameters $a$ and $b$ is defined as continuous within a finite interval. Therefore, the output layers of the neural network are designed to have two channels for each parameter, and the conditional probabilities at each site are expressed as \cite{Germain:2015yft}
\begin{eqnarray}
    f_{\theta}(s_{i}|s_{1}, \cdots, s_{i- 1})=\frac{\Gamma(a_{i}+b_{i})}{\Gamma(a_{i})\Gamma(b_{i})}s_{i}^{a_{i}-1}(1-s_{i})^{b_{i}-1},
\end{eqnarray}
where $\Gamma(a)$ is the gamma function, $\{\theta\}$ is a set of trainable parameters of the networks, and $s_i$ is the configuration of the system. The outputs of the hidden layers are $a \equiv (a1, a2, ...)$ and $b \equiv (b1, b2, ... )$. The conditional probability is realized by adding a mask which veils all the sites whose indices are larger than a given position before one sample is conveyed to convolutional       layers. This setup agrees with the locality of microscopic interactions and is capable of preserving any high-order interactions from a restricted Boltzmann machine perspective~\cite{Aarts:2023uwt}. With such a conditional probability \textit{Ansatz}, the joint probability of a sample is as
\begin{eqnarray}
    q_{\theta}(s)=\prod_{i = 1}^{N}f_{\theta}(s_{i}|s_{1}, \cdots, s_{i-1}).
\end{eqnarray}
The loss function in training is designed by maximize the probability of the ensemble (training set) according to the MLE principle, i.e. the most physical ensemble is the most possible to emerge. Hence the loss is calculated as the logarithmic value of the mixture distribution obtained from the network, 
\begin{eqnarray}
    L=-\sum_{s\sim q_(data)}\log(q_{\theta}(s))
\end{eqnarray}
and Adam optimizer is used to minimize this loss. With this framework the problem of density estimation is converted to find the $\{a_i\}$ and  $\{b_i\}$ for each site which are working for all the samples in the ensemble. In this way we have sidestepped finding different probability(unknown) for each sample and achieved the positivity and similarity of the sample probability. The only \textit{Ansatz} is the way of factorizing the whole probability of a sample. It would work better if the explicit form of the conditional probability (Beta distribution in this work) is chosen to have better expression ability. 

In an unsupervised manner, the training process is equivalent to find the correct probability of each sample. The training data we need is only two physically distributed ensembles at different temperatures and no analytical Lagrangian density expression are required by the CAN. These training data are obtained via the traditional MCMC simulation \cite{Lopez-Ruiz:2016tnf}.  

\subsection{Action Estimation}
\begin{figure}[!hbtp]
\centering
\includegraphics[width=0.5\textwidth]{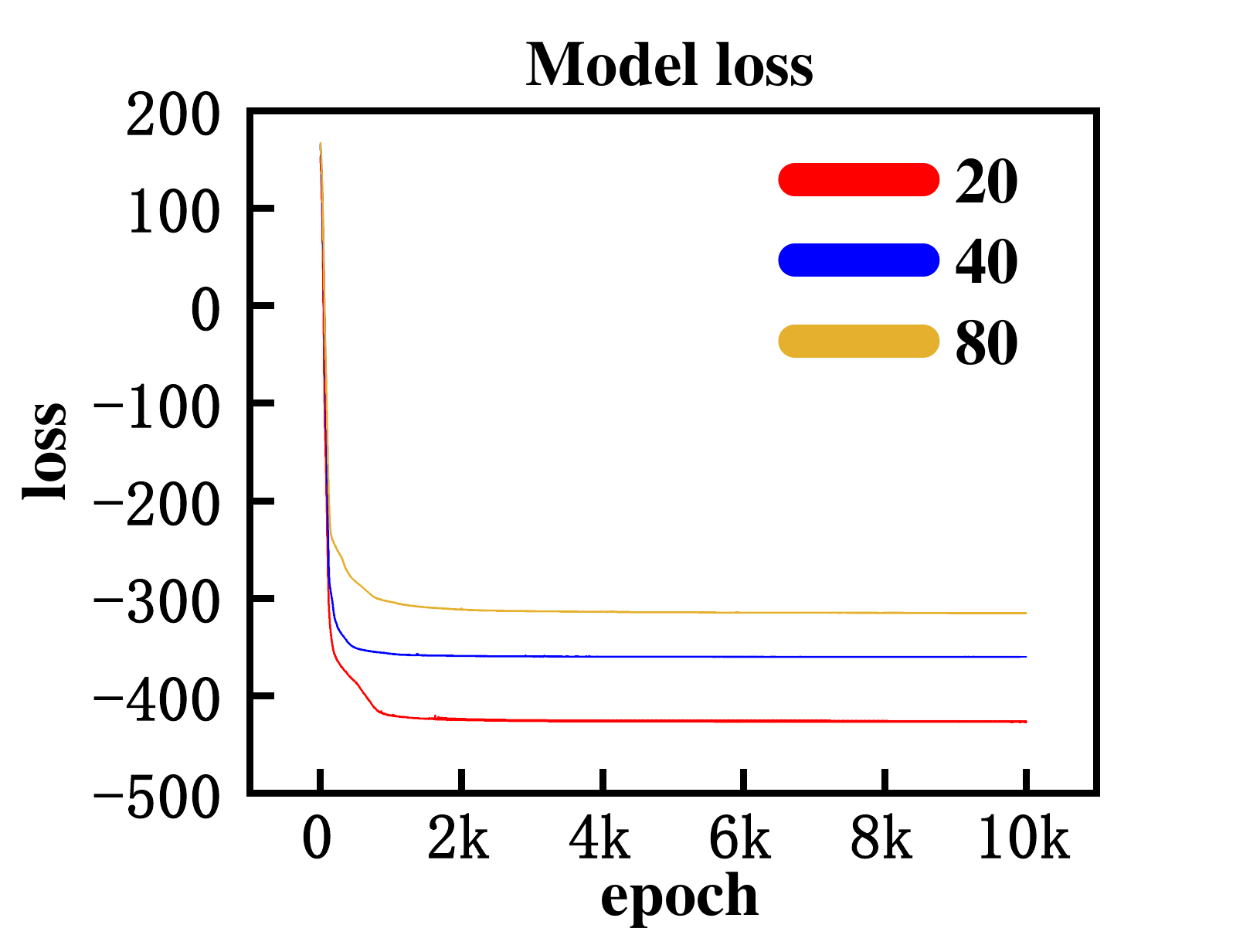}
\caption{\label{loss}Model loss for $\beta=$20, 40, 80, respectively}
\end{figure}

In this section we show the training and validation stages of the CANs. It will be seen that the CANs can succeed to reproduce the correct action of most configurations at different temperature without knowing the analytical form of the Lagrangian density (unsupervisely).
To train the network we have prepared three ensembles at $\beta=T^{-1}=$20, 40 and 80 with the MCMC simulation. We train the CANs to estimate the probability of each sample in these three ensembles for 10,000 epochs. { From Figure~\ref{loss} we can find the $\beta=40$ case converges fastest, while the $\beta=80$ case is slower and $\beta=80$ slowest. This means the network is good at learning the ensemble with different kinds of configurations, because both the kinetic and potential terms will contribute to the averaged loss equally. Two limit cases is harder to learn because the network will focus on the kinetic/potential part. As a result the loss from the other part will be more difficult to reduce. Nevertheless with a large enough number of epoch(larger than 2000) all the networks has been convergent.}

\begin{figure}
\centering
\includegraphics[width=0.5\textwidth]{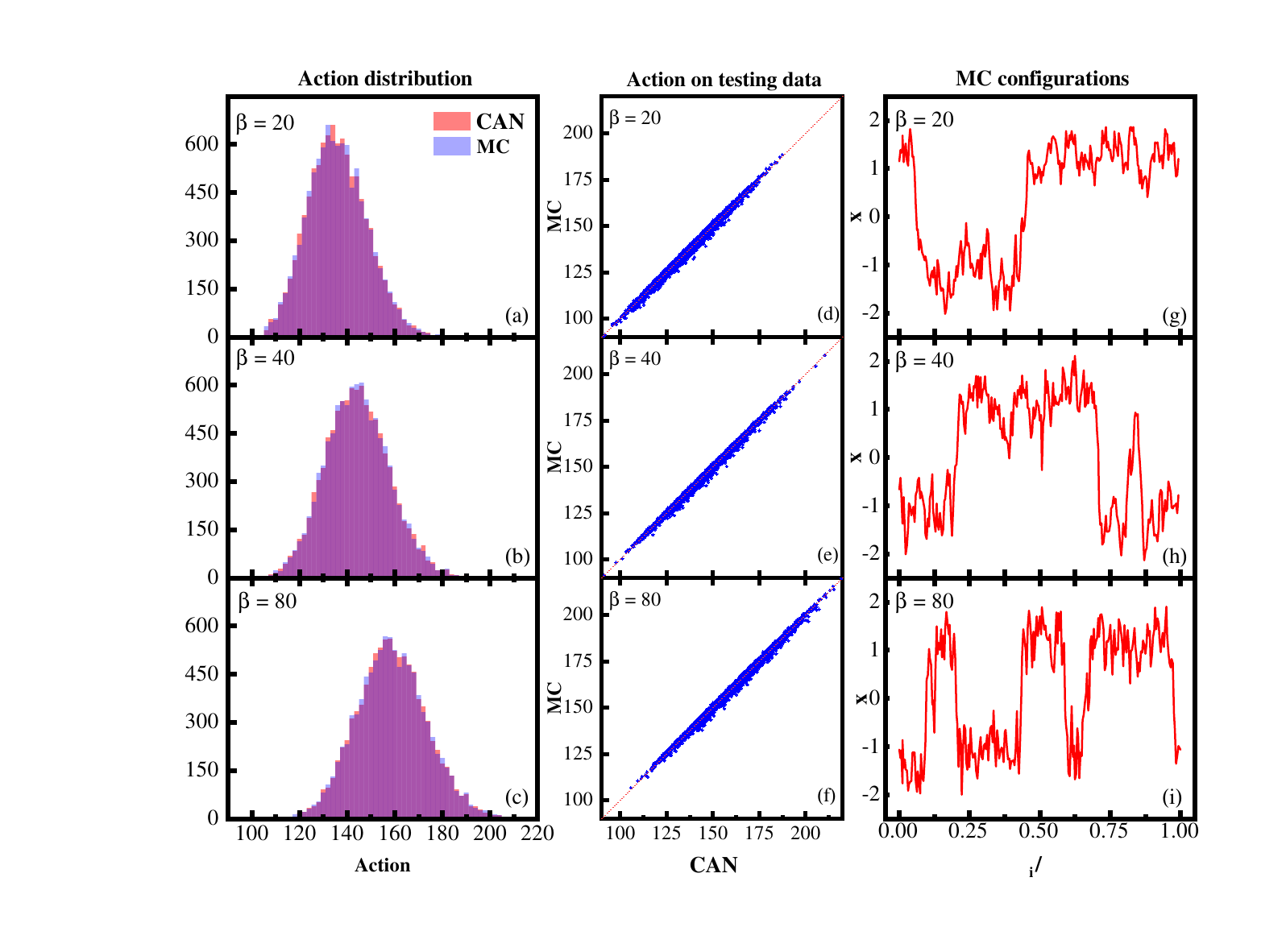}
\caption{\label{onebeta}(a)-(c) Comparisons of action distribution by MCMC(analytical) and CANs; (d)-(f) Comparisons of action on testing data by MC and CANs; (g)-(i)Monte Carlo configuration ${x_i}$ for $\beta=$20, 40, 80, respectively}
\end{figure}

The CANs can estimate the probability well as shown in Figure ~\ref{onebeta}. The first column shows the histograms of both the analytical (blue) and CANs-output (red) actions of the test data. Because the constant in Eq.~\ref{interp} can not be determined, we have shifted the histogram of CAN by hand to make their peaks at the same position. And the same shifting constant is adopted in the further comparison. It is clear that they almost coincide with each other. To show the estimating ability, a sample-wise comparison of these two actions are shown in the second column. Up to the above-mentioned constant the two actions fall on a straight line whose slope equals 1. This means the CANs can not only reproduce the correct distribution (histogram) of action for a given ensemble but more important the correct action for each sample. And the third column show typical configurations in the ensembles at each temperature. At larger $\beta$(low temperature) the configuration appears to be a link of more multi kink and anti kinks. When temperature increases, the number of kinks and anti-kinks will reduce and vanishes eventually. This behavior agrees with our expectation and means the ensembles distribute correctly. From the results it is shown clearly that the CANs have the capacity to learn the possibility distribution at one single temperature and successfully learned most of the samples' action up to a network-dependent constant without introducing the analytical Lagrangian density into the CANs. 

Once two networks, corresponding two different temperatures, have been trained, the action of each sample in the third ensemble at a different temperature can be extracted with Eq.\ref{interp} up to a sample-independent but network dependent constant. In next section we will use networks trained at two temperature of the above three cases to predict the sample actions at the third temperature.

\subsection{Action Prediction}
\begin{figure}[!hbpt]
\centering
\includegraphics[width=0.5\textwidth]{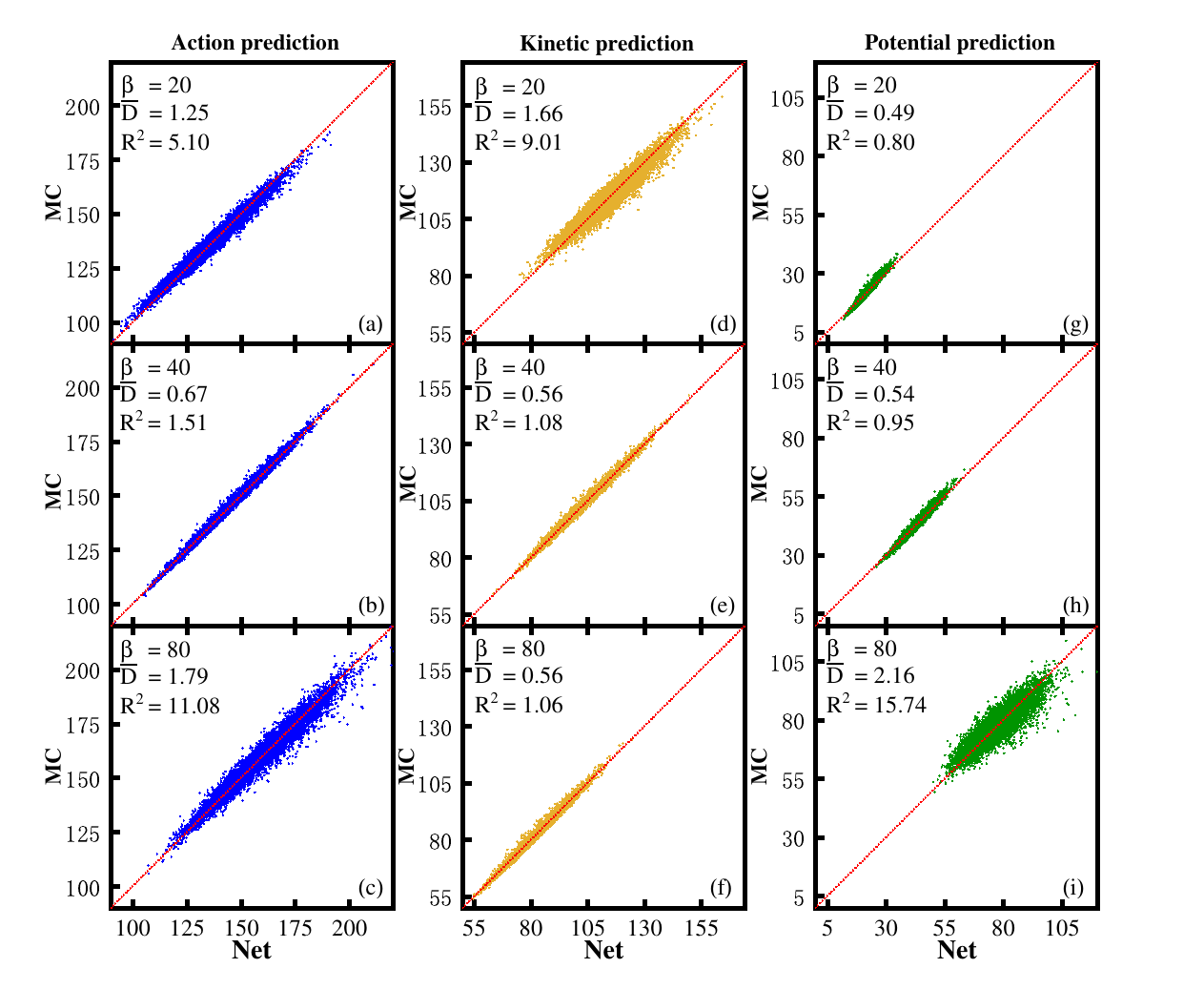}
\caption{\label{twobeta}(a)-(c)Comparisons of action on testing data by MC and prediction Net; (d)-(f) The same comparison for the kinetic part. (g)-(i) The same comparison for the potential part.}
\end{figure}

With the trained networks it is ready for estimation actions from those at two other temperatures. By pretending having neither the network nor the analytical Lagrangian density at $\beta_3$, we start to find the action of an arbitrary sample at $\beta_3$ by making use of two trained neural networks at $\beta_1$ and $\beta_2$ according to the procedure presented in Section~\ref{sec:quantum} using Eq.\ref{interp}. As we have already obtained three networks at three temperatures, we will choose any two of them as $\beta_1$ and $\beta_2$ to predict the remaining one $\beta_3$ whose network becomes idle. The predicted total action, kinetic part (proportional to $\beta^{-1}$) and  potential part (proportional to $\beta$) of the 2-to-1 prediction and analytical actions are show in the first, second and third column respectively in { Figure~\ref{twobeta}}. Clearly the first ($\beta_{40, 80}$ to $\beta_{20}$) and last ($\beta_{20, 40}$ to $\beta_{80}$) rows are extrapolation cases and the second ($\beta_{20, 80}$ to $\beta_{40}$) is the interpolation. { In these figures we list two quantities, i.e. $R^2$ and $\bar{D}$ to show the predicting ability of the CANS. $R^2$ is the mean value of the square of the predicting error $R^2=\langle{(S_{True}-S_{CANs})^2}\rangle$ and the $\bar{D}$ is the mean of the distance to red line with the slope 1.} We can find that the interpolation case, i.e. the $\beta_{20, 80}$ to $\beta_{40}$, works best. 

The reason can be concluded if we notice the kinetic and potential part distributions. In the  $\beta_{40, 80}$ to $\beta_{20}$ case (first row), the $\beta_{40}$ and $\beta_{80}$ ensembles are more dominated by multi-kink and anti-kink configurations whose kinetic parts are relatively larger because of the larger derivative from the jump, i.e., $\pm 2$ to $\mp 2$ in our computation. However predicted ensembles  $\beta_{20}$ have less kinks/anti-kinks, this makes the kinetic part estimation is not good enough. On the other hand, in the $\beta_{20, 40}$ to $\beta_{80}$ case (third row), the $\beta_{80}$ ensemble has more multi-kinks/anti-kinks, i.e. more sites value equals $\pm 2$, while in $\beta_{20, 40}$ ensembles more sites absolute value less than 2. This makes the potential part estimation not good enough. In the interpolation case, the training data includes ensembles both at low and high temperature, which makes the training data cover more different configurations, so the kinetic and potential parts both work well in the $\beta_{20, 80}$ to $\beta_{40}$ case. Physically, the low and high temperature ensembles typically corresponds distinct phases of the system. Once the network has assimilated the information, it is equipped to predict the system's behavior at any intermediate temperature stage. This capability is crucial for detailed exploration of the phase diagram.

\section{Conclusion}
In this work, we propose a paradigm for constructing an effective model using Artificial Neural Networks (ANNs) once the ensemble of a certain degree of freedom (\textit{d.o.f}) has been obtained. Utilizing Continuous Autoregressive Networks (CANs) and a Higgs-like 0+1D quantum field model, we demonstrate the construction process. For this model, there is a topological phase transition from a state dominated by kinks and anti-kinks to a state without kinks as the temperature increases from low to high. Using ensembles generated by traditional Markov Chain Monte Carlo (MCMC), the CANs successfully extract the probability of each sample. By utilizing the two trained networks at different temperatures, the action of a sample at an arbitrary third temperature can be easily determined using Equation~\ref{interp}. As expected, the predictions are most accurate when interpolating. This approach is beneficial for constructing a new effective model targeting specific degrees of freedom (\textit{d.o.f}), once the ensembles of the fundamental \textit{d.o.f} have been established.

This novel paradigm is particularly effective for investigating phase transitions e.g., deconfinement, distinguishing it from other applications of supervised learning applications~\cite{Boyda:2020nfh,Favoni:2020reg,Palermo:2021jrf}. Additionally, it is more user-friendly compared to existing unsupervised methods~\cite{Sale:2022qfn,Spitz:2022tul}. For instance, consider Quantum Chromodynamics (QCD) at finite temperature. It is suggested that around the critical temperature for deconfinement, a soliton solution of gluons, known as a dyon, dominates the system.
With complicated computation, the gluon-quark system governed by QCD can be converted into a dyon-quark ensemble, wherein dyons interact in intricate ways. Although such a dyon ensemble has been obtained analytically after extensive efforts \cite{Diakonov:2004jn}, further calculations regarding physical observables would still rely on numerical simulation. Consequently, employing the methodology described in this work will be advantageous for developing an effective model using Artificial Neural Networks (ANNs). 

If the ensemble of gluon configurations can be obtained through lattice simulation, it is possible to first transform the fundamental gluon field into multiple dyons and anti-dyons in space, thereby converting the gluon ensemble into a dyon ensemble. Subsequently, an effective numerical model can be developed by utilizing these dyon ensembles with the methodology described in this work. As suggested above, the numerical model comprises two trained networks. At any given temperature, the action at the third temperature can be calculated. This action can then be utilized as if it were derived from the analytical dyon ensemble model. The same procedure can be applied to any effective degrees of freedom (\textit{d.o.f}) of various systems.

\section*{Acknowledgements}
The work of this research is supported by the National Natural Science Foundation of China, Grant Nos. 12375131(YJ), 12375136(LH), the CUHK-Shenzhen university development fund under grant No. UDF01003041 and the BMBF funded KISS consortium (05D23RI1) in the ErUM-Data action plan (KZ).
L. Wang also thanks the National Natural Science Foundation of China[Grant No.12147101] for supporting his visit to Fudan University.

\end{document}